\documentclass[twocolumn,english,aps,prd,reprint,floatfix,notitlepage,footinbib,preprintnumbers,superscriptaddress,longbibliography]{revtex4-1}
\pdfoutput=1
\usepackage{lmodern}

\usepackage[T1]{fontenc}
\usepackage[latin9]{inputenc}
\usepackage{geometry}
\usepackage{subfigure,lmodern, amsmath,amssymb, graphicx, pifont, adjustbox, bm, xcolor}
\usepackage{amsfonts}
\usepackage{enumitem}
\usepackage{comment}
\usepackage{mathtools}
\usepackage{float}
\usepackage{slashed}
\usepackage{ragged2e}
\usepackage{array}
\usepackage{microtype}
\usepackage{soul}

\usepackage{nameref}

\usepackage{hhline}

\makeatletter\g@addto@macro\bfseries{\boldmath}\makeatother

\makeatletter\newcommand{\labeltext}[2]{%
  \def\@currentlabel{#1}%
  \label{#2}%
}
\makeatother

\usepackage{stackengine}
\usepackage{esint}
\usepackage[unicode=true,pdfusetitle,
 bookmarks=true,bookmarksnumbered=false,bookmarksopen=false,
 breaklinks=false,pdfborder={0 0 1},backref=false,colorlinks=true]
 {hyperref}
\hypersetup{
 pdfauthor={Clifford Cheung, Jaehoon Jeong, Pyungwon Ko, Alex Pomarol, Grant N. Remmen, Francesco Sciotti},
 citecolor=black,linkcolor=black,urlcolor=black}

\newcommand{\appendixref}[1]{\hyperref[#1]{appendix~\ref{#1}}}
\def\equationautorefname~#1\null{eq.\,(#1)\null}
\usepackage{breakurl}
\usepackage[hang,flushmargin]{footmisc} 
\allowdisplaybreaks
\makeatletter

\usepackage{etoolbox}
\apptocmd{\thebibliography}{\justifying\setlength{\leftskip}{7.4mm}}{}{} 
 
 \usepackage{relsize}
\usepackage{babel}

\usepackage{bbm}

\makeatletter
\def\simgt{\mathrel{\lower2.5pt\vbox{\lineskip=0pt\baselineskip=0pt
           \hbox{$>$}\hbox{$\sim$}}}}
\def\simlt{\mathrel{\lower2.5pt\vbox{\lineskip=0pt\baselineskip=0pt
           \hbox{$<$}\hbox{$\sim$}}}}
\makeatother

\usepackage{changepage}

\newcommand{\be}{\begin{equation}}
\newcommand{\ee}{\end{equation}}
\newcommand{\bea}{\begin{eqnarray}}
\newcommand{\eea}{\end{eqnarray}}

\newcommand{\Eq}[1]{Eq.~(\ref{#1})}

\newcommand{\eq}[2]{\be\begin{aligned}#1 \label{#2}\end{aligned}\ee}
\newcommand{\mysec}[1]{\noindent {\bf #1.}---}

\newcommand{\cdotn}{\cdot\!\cdot\!\cdot}

\newcommand{\Fpi}{F_\pi}

\newcolumntype{P}[1]{>{\centering\arraybackslash}p{#1}}

\usepackage{scalerel}

\usepackage{fix-cm}

\definecolor{dartmouthgreen}{rgb}{0.05, 0.5, 0.06}

\usepackage{tikz}
\usetikzlibrary{calc}
\newcommand{\blobdiam}{8mm}      % change this to resize all blobs
\colorlet{Lfill}{blue!12}        % color of F^(L) blobs
\colorlet{Rfill}{orange!12}      % color of F^(R) blobs

\tikzset{
  Lblob/.style={
    circle,
    draw,
    fill=Lfill,
    line width=0.9pt,
    minimum size=\blobdiam,
    inner sep=0pt
  },
  Rblob/.style={
    circle,
    draw,
    fill=Rfill,
    line width=0.9pt,
    minimum size=\blobdiam,
    inner sep=0pt
  },
  leg/.style={line width=0.9pt},
  prop/.style={line width=0.9pt},
  mom/.style={font=\scriptsize, inner sep=1pt},
  bloblab/.style={font=\scriptsize}
}

\newcommand{\FiveMixed}[3]{%
\begin{scope}[shift={({#2},{#3})}]
  \node[Lblob] (#1L) at (0,0) {};
  \node[Rblob] (#1R) at (1.85,0) {};
  \draw[prop,font=\large] (#1L.east) -- node[above, font=\large] {$k$} (#1R.west);

  \coordinate (#1p1) at (-0.95,  0.55);
  \coordinate (#1p2) at (-0.95, -0.55);
  \coordinate (#1p3) at ( 2.75,  0.55);
  \coordinate (#1p4) at ( 3.00,  0.00);
  \coordinate (#1p5) at ( 2.75, -0.55);

  \draw[leg] (#1p1) -- (#1L.145);
  \draw[leg] (#1p2) -- (#1L.215);
  \draw[leg] (#1R.35) -- (#1p3);
  \draw[leg] (#1R.east) -- (#1p4);
  \draw[leg] (#1R.-35) -- (#1p5);

  \node[bloblab,font=\large] at ($(#1L.south)+(0,-0.42)$) {$A_k^{(2+1)}$};
  \node[bloblab,font=\large] at ($(#1R.south)+(0,-0.42)$) {$A_k^{(3+1)}$};
\end{scope}
}

\newcommand{\FourDiag}[3]{%
\begin{scope}[shift={({#2},{#3})}]
  \node[Lblob] (#1L) at (0,0) {};
  \node[Lblob] (#1R) at (1.85,0) {};
  \draw[prop,font=\large] (#1L.east) -- node[above, font=\large] {$k$} (#1R.west);

  \coordinate (#1p1) at (-0.95,  0.55);
  \coordinate (#1p2) at (-0.95, -0.55);
  \coordinate (#1p3) at ( 2.80,  0.55);
  \coordinate (#1p4) at ( 2.80, -0.55);

  \draw[leg] (#1p1) -- (#1L.145);
  \draw[leg] (#1p2) -- (#1L.215);
  \draw[leg] (#1R.30) -- (#1p3);
  \draw[leg] (#1R.-30) -- (#1p4);

  \node[bloblab,font=\large] at ($(#1L.south)+(0,-0.42)$) {$A_k^{(2+1)}$};
  \node[bloblab,font=\large] at ($(#1R.south)+(0,-0.42)$) {$\bar{A}_k^{(2+1)}$};
\end{scope}
}

\newcommand{\SixDiag}[3]{%
\begin{scope}[shift={({#2},{#3})}]
  \node[Rblob] (#1L) at (0,0) {};
  \node[Rblob] (#1R) at (1.95,0) {};
  \draw[prop] (#1L.east) -- node[above, font=\large] {$k$} (#1R.west);

  \coordinate (#1p1) at (-1.05,  0.70);
  \coordinate (#1p2) at (-1.20,  0.00);
  \coordinate (#1p3) at (-1.05, -0.70);

  \coordinate (#1p4) at ( 3.00,  0.70);
  \coordinate (#1p5) at ( 3.15,  0.00);
  \coordinate (#1p6) at ( 3.00, -0.70);

  \draw[leg] (#1p1) -- (#1L.150);
  \draw[leg] (#1p2) -- (#1L.west);
  \draw[leg] (#1p3) -- (#1L.210);

  \draw[leg] (#1R.30) -- (#1p4);
  \draw[leg] (#1R.east) -- (#1p5);
  \draw[leg] (#1R.-30) -- (#1p6);

  \node[bloblab,font=\large] at ($(#1L.south)+(0,-0.42)$) {$\bar{A}_k^{(3+1)}$};
  \node[bloblab,font=\large] at ($(#1R.south)+(0,-0.42)$) {$A_k^{(3+1)}$};
\end{scope}
}

\begin{document}

\preprint{CALT-TH 2026-019}
\preprint{KIAS-Q26009}

\title{Multipositivity Constrains the Chiral Lagrangian}

\author{Clifford Cheung}
\affiliation{Walter Burke Institute for Theoretical Physics and
Leinweber Forum for Theoretical Physics, California Institute of Technology, Pasadena, CA 91125, USA}
\author{Jaehoon Jeong}
\affiliation{\scalebox{1}{Quantum Universe Center, KIAS, Seoul 02455, Korea}}
\author{Pyungwon Ko}
\affiliation{\scalebox{1}{Quantum Universe Center, KIAS, Seoul 02455, Korea}}
\author{Alex Pomarol}
\affiliation{\scalebox{1}{IFAE, BIST and Dept. de F\'isica, Universitat Aut\`onoma de Barcelona, 08193 Bellaterra, Barcelona, Spain}}
\author{Grant N.~Remmen}
\affiliation{\scalebox{1}{Center for Cosmology and Particle Physics, Department of Physics, New York University, New York, NY 10003, USA}}    
\author{Francesco Sciotti}
\affiliation{\scalebox{1}{IFAE, BIST and Dept. de F\'isica, Universitat Aut\`onoma de Barcelona, 08193 Bellaterra, Barcelona, Spain}}
    
\begin{abstract} 

\noindent The chiral Lagrangian is a cornerstone of modern particle physics, offering a systematic and quantitative description of low-energy pions. Using tools from the modern scattering amplitudes program, we show that consistent multiparticle dynamics impose novel constraints on the coupling constants of this theory. In the planar limit, these constraints imply that certain Wilson coefficients of the chiral Lagrangian are bounded from below by the chiral anomaly. Our results reveal a subtle connection between the anomalous and nonanomalous sectors of the underlying strong interactions, while introducing a novel formulation of multipositivity bounds that holds for any planar tree-level theory.

\noindent 
\end{abstract}
\maketitle 

\mysec{Introduction}The advent of chiral perturbation theory revolutionized our understanding of the strong interactions.  In a single stroke, it reforged the insights of current algebra~\cite{Gell-Mann:1960mvl} into the first incarnation of effective field theory (EFT)~\cite{Weinberg:1966fm,Cronin:1967jq,Weinberg:1968de,Coleman:1969sm,Callan:1969sn,Weinberg:1978kz,Gasser:1984gg}, in modern parlance known as the nonlinear sigma model (NLSM).  This framework trades the intricate short-distance physics of quantum chromodynamics (QCD) for {\it parameters} in a long-distance description dictated by symmetry and dimensional analysis. More than half a century later, chiral perturbation theory remains one of the most celebrated, well-studied, and quantitatively successful EFTs in all of physics.

Beyond its phenomenological roots, the NLSM has more recently been a central object of study in the more formal context of the modern scattering amplitudes program.
Indeed, NLSM amplitudes are among the handful of models---along with gauge theory and general relativity---that exhibit numerous extraordinary mathematical structures, including soft theorems~\cite{Susskind:1970gf,Osborn:1969ku,Cheung:2014dqa,Low:2019ynd,Cheung:2016drk,Kampf:2012fn,Kampf:2013vha,Bartsch:2024ofb}, double copy structures~\cite{Cachazo:2014xea,Cachazo:2013iea,Chen:2013fya,Carrasco:2016ygv,Carrasco:2016ldy,Cheung:2016prv,Cheung:2017yef,Cheung:2017ems}, 
and remarkable representations for scattering in terms of abstract volumes~\cite{Arkani-Hamed:2017mur,Arkani-Hamed:2023swr,Arkani-Hamed:2024nhp,Arkani-Hamed:2024yvu}.

Given the foundational status of the chiral Lagrangian, one might expect its parameter space to be well-charted territory. Curiously, even decades after its inception, it is not fully understood what is the allowed space of consistent coupling constants, even in principle.
In recent years a program has been initiated to systematically answer this question using consistency constraints---or so-called {\it positivity bounds}---which can be derived solely from general principles like unitarity and causality~\cite{Adams:2006sv}. Positivity bounds have shown themselves to be a powerful tool to constrain EFTs~\cite{Bellazzini:2020cot,Arkani-Hamed:2020blm,Caron-Huot:2020cmc}, with applications ranging from the Standard Model to quantum gravity~\cite{Albert:2022oes, Fernandez:2022kzi, Albert:2023jtd, Ma:2023vgc, Albert:2023seb, Li:2023qzs, Dong:2024omo, Dong:2025dpy,Nicolis:2009qm,Tolley:2020gtv,Pham:1985cr,Ananthanarayan:1994hf,Pennington:1994kc,Comellas:1995hq,Dita:1998mh,Manohar:2008tc,Mateu:2008gv,Jenkins:2006ia,Dvali:2012zc,completeness,Arkani-Hamed:2021ajd,Cheung:2018cwt,Cheung:2019cwi,EliasMiro:2022xaa,Caron-Huot:2021rmr,Bellazzini:2025shd,Bellazzini:2015cra,Cheung:2016wjt,Camanho:2014apa,Gruzinov:2006ie,Arkani-Hamed:2021ajd,Cheung:2018cwt,Cheung:2019cwi,Cheung:2014ega,Bellazzini:2019xts,Caron-Huot:2022ugt,Caron-Huot:2022jli,Cheung:2016yqr,deRham:2017xox,Bellazzini:2023nqj,Bern:2021ppb,Berman:2023jys,Freytsis:2022aho,Remmen:2019cyz,Remmen:2020vts,Remmen:2020uze,Bellazzini:2016xrt,Remmen:2022orj,Remmen:2024hry,Bi:2019phv,Zhang:2018shp,Berman:2024wyt,Remmen:2021zmc,Bachu:2022gof,Huang:2022mdb,Cheung:2022mkw,Cheung:2023adk,Cheung:2023uwn,Haring:2023zwu,Haring:2024wyz,Arkani-Hamed:2022gsa,Bhardwaj:2024klc,Aoude:2024xpx,SFAN,Guerrieri:2021ivu,Albert:2024yap,Berman:2024eid,Elvang:2026pmc,Cheung:2024uhn,Cheung:2024obl,Cheung:2025nhw,Arkani-Hamed:2023jwn,Basile:2026gnd,Chandrasekaran:2018qmx,Berman:2025owb,Saha:2026ftv,Bresciani:2025toe,Hillman:2024ouy,Caron-Huot:2016icg,Bellazzini:2025bay,Huang:2025icl,Beadle:2025cdx,Beadle:2024hqg,Pasiecznik:2025eqc,Caron-Huot:2024lbf,He:2023lyy,Cordova:2019lot,Guerrieri:2018uew,Guerrieri:2020bto,EliasMiro:2026kww,EliasMiro:2023fqi,EliasMiro:2022xaa,Eckner:2024ggx,Eckner:2024pqt}. 

Most existing results rely on exploiting positivity properties of two-to-two scattering amplitudes. Recent work has begun to extend these ideas to higher multiplicity~\cite{Cheung:2025nhw,Arkani-Hamed:2023jwn,Bresciani:2025toe,Chandrasekaran:2018qmx,Berman:2025owb,Saha:2026ftv,Basile:2026gnd}, with Ref.~\cite{Cheung:2025nhw} introducing forms of {\it multipositivity} that relate amplitudes with different numbers of external particles. The methods introduced in this paper build on this direction. We focus on dispersion relations in a single kinematic channel and exploit their factorization structure, allowing us to apply a Cauchy-Schwarz type of inequality directly to residues. This construction leads to a new class of bounds that relate planar amplitudes of different multiplicities in a simple and controlled manner. 

This approach is particularly well suited to the chiral Lagrangian, where symmetry relates interactions across different pion multiplicities. As we will show, we find new constraints on chiral EFT data that are invisible to four-point scattering and that carve out a significant portion of the previously allowed parameter space. 

\medskip

\mysec{Planar Amplitudes in Large-$N_c$ QCD}In this paper, we derive new constraints on higher-point \emph{planar} amplitudes. 
The planar limit is particularly convenient because the singularity structure of the amplitude simplifies, as will become clear below.
As a concrete and physically relevant arena for these ideas, we will focus on large-$N_c$ QCD, an $SU(N_c)$ gauge theory in the $N_c\to \infty$ limit~\cite{tHooft:1973alw,Witten:1979kh}, with $N_f$ massless quarks in the fundamental representation. The theory exhibits the familiar pattern of chiral symmetry breaking, $U(N_f)_L \times U(N_f)_R \to U(N_f)_V$.  The massless spectrum consists of Goldstone bosons in the adjoint of $SU(N_f)$, the pions, together with the singlet $\eta^\prime$. In the large-$N_c$ limit, the strong dynamics of quarks and gluons enjoy a dual description in terms of the tree-level scattering of weakly interacting mesons. 

At leading order in the large-$N_c$ expansion, connected pion amplitudes are generated by planar diagrams with a \emph{single} quark boundary~\cite{tHooft:1973alw,Witten:1979kh}. Since the quarks are in the fundamental representation, this boundary line carries flavor indices. The only structures that can appear in flavor space are therefore single traces~\footnote{Multi-trace flavor structures require additional quark boundaries (equivalently, extra disconnected flavor loops) and are therefore suppressed by further powers of $1/N_c$.}. In other words, though $N_f$ is finite, the large-$N_c$ expansion organizes pion amplitudes into single-trace flavor structures that define flavor-ordered partial amplitudes,
\begin{equation}
A^{a_1\cdots a_n} \!= \!\sum_{\sigma \in S_n/\mathbb{Z}_n} \!\! \!\text{Tr}^{a_{\sigma(1)}\cdots a_{\sigma(n)}} A^{(n)}(p_{\sigma(1)},\!...,p_{\sigma(n)}),
\label{nptplanar}
\end{equation}
where the momenta $p_i$ are associated to the pions of flavor $a_i$.  Here $\text{Tr}^{a_1\cdots a_n}=\text{Tr}(T^{a_1}\cdotn T^{a_n})$ in the normalization convention $\text{Tr}(T^a T^b) = \delta^{ab}$, which is required for planar amplitudes to factorize correctly.   Cyclicity of the traces implies that the partial amplitudes are also cyclic, $A^{(n)}(p_1,p_2,\!...,p_n)=A^{(n)}(p_2,\!...,p_n,p_1)$,
which at four-point is nothing more than the $s\leftrightarrow t$ crossing symmetry we are accustomed to.
A  key property of tree-level pion amplitudes  at large $N_c$  is that their
singularities correspond to simple poles, associated to the exchanged mesons, that show up {\it only} in planar channels. 
The planar partial amplitudes $A^{(n)}$ will be the central objects of study in our higher-point analysis. 

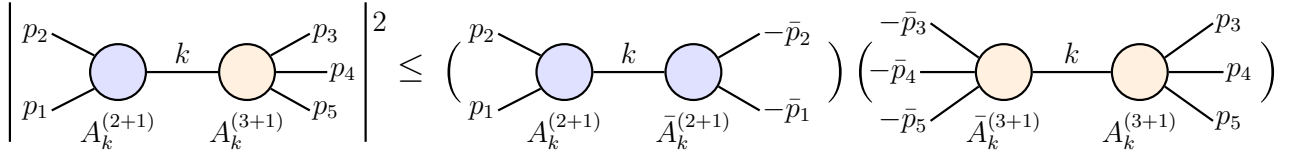
\begin{figure*}[t]
\centering
\resizebox{0.95\textwidth}{!}{%
\begin{tikzpicture}[baseline={(current bounding box.center)}]

  % --------------------------------
  % Left-hand side
  % --------------------------------
  \FiveMixed{LHS}{0}{0}

  % absolute value bars and square
  \draw[line width=0.9pt] (-1.55, -1.00) -- (-1.55, 1.00);
  \draw[line width=0.9pt] ( 3.55, -1.00) -- ( 3.55, 1.00);
  \node[font=\large] at (3.75,0.72) {$2$};

  % inequality sign
  \node[font=\Large] at (4.2, 0.05) {$\leq$};

  % --------------------------------
  % First factor on RHS
  % --------------------------------
  \node[font=\Huge] at (4.8,0.00) {$($};
  \FourDiag{RHSA}{6.4}{0}
  \node[font=\Huge] at (10.3,0.00) {$)$};

  % --------------------------------
  % Second factor on RHS
  % --------------------------------
  \node[font=\Huge] at (10.7,0.00) {$($};
  \SixDiag{RHSB}{12.7}{0}
  \node[font=\Huge] at (16.5,0.00) {$)$};

  % --------------------------------
  % Example momentum labels
  % (edit these freely later)
  % --------------------------------

  % LHS
  \node[mom, left,font=\large]  at (LHSp1) {$p_2$};
  \node[mom, left,font=\large]  at (LHSp2) {$p_1$};
  \node[mom, right,font=\large] at (LHSp3) {$p_3$};
  \node[mom, right,font=\large] at (LHSp4) {$p_4$};
  \node[mom, right,font=\large] at (LHSp5) {$p_5$};

  % 4pt diagonal
  \node[mom, left,font=\large]  at (RHSAp1) {$p_2$};
  \node[mom, left,font=\large]  at (RHSAp2) {$p_1$};
  \node[mom, right,font=\large] at (RHSAp3) {$-\bar{p}_2$};
  \node[mom, right,font=\large] at (RHSAp4) {$-\bar{p}_1$};

  % 6pt diagonal
  \node[mom, left,font=\large]  at (RHSBp1) {$-\bar{p}_3$};
  \node[mom, left,font=\large]  at (RHSBp2) {$-\bar{p}_4$};
  \node[mom, left,font=\large]  at (RHSBp3) {$-\bar{p}_5$};

  \node[mom, right,font=\large] at (RHSBp4) {$p_3$};
  \node[mom, right,font=\large] at (RHSBp5) {$p_4$};
  \node[mom, right,font=\large] at (RHSBp6) {$p_5$};

\end{tikzpicture}%
}
\caption{Inequality satisfied by the residues of the four-, five- and six-point scattering amplitudes, which leads to \Eq{eq:CS-res}. 
The state sum on the cut runs over exchanged particles labeled by $k$.}
\label{fig:456bound}
\end{figure*}

\medskip
\mysec{Chiral Perturbation Theory}In chiral perturbation theory, the parity-even, nonanomalous dynamics are encoded by the NLSM Lagrangian in terms of $U=\exp(\sqrt{2}i \pi^a T^a / F_\pi )$, which transforms in the bifundamental 
of $U(N_f )_L \times U( N_f )_R$:
\begin{equation}
\begin{aligned}
\mathcal{L}_{\rm NLSM} =&\, -\frac{F_\pi^2}{4} \,\text{Tr}(\partial_\mu U^\dagger \partial^\mu U) \\&\, +   L_3 \text{Tr}(\partial_\mu U^\dagger\partial^\mu U\partial_\nu U^\dagger \partial^\nu U) \\&\,+L_4 \text{Tr}(\partial_\mu U^\dagger\partial_\nu U\partial^\mu U^\dagger \partial^\nu U) + \cdots ,
\label{chirallagrangian}
\end{aligned}
\end{equation}
where  we work in mostly-plus signature.  We will ignore higher-trace operators, which are subleading at large $N_c$, as well as any additional gauge interactions. 

Due to the chiral anomaly,  there is an additional Wess-Zumino-Witten (WZW) term~\cite{Wess:1971yu,Witten:1983tw}, which breaks $U \to U^\dagger$ invariance while preserving parity, which sends $U\to U^\dagger$ and $\vec{x}\to-\vec{x}$. This topological term is constructed as an integral over a five-dimensional manifold, with physical four-dimensional spacetime as its boundary. 
Defining the chiral current $j_\mu = i\, U^\dagger \partial_\mu U$, the leading intrinsic parity-odd, anomalous dynamics come from
\eq{
S_{\text{WZW}} = -\frac{\kappa}{240 \pi^2} \int d^5x\, \epsilon^{\mu\nu\rho\sigma\tau} \text{Tr} \left( j_\mu j_\nu j_\rho j_\sigma j_\tau \right),
}{WZWpion}
from which Stokes' theorem yields (if $N_f\geq3$) the anomalous five-point vertex in four dimensions,
\be\hspace{-1mm}
{\cal L}_{\rm WZW} = \frac{\kappa\, \text{Tr}^{abcde}\epsilon^{\mu\nu\rho\sigma} \pi^a \partial_\mu \pi^b \partial_\nu \pi^c \partial_\rho \pi^d \partial_\sigma \pi^e}{30\sqrt{2} \pi^2 \Fpi^5}  {+} \cdots. 
\ee
Matching to the QCD chiral anomaly~\cite{Witten:1983tw} precisely fixes
$\kappa = N_c$.
Since $S_{\rm WZW}$ is odd under $U \to U^\dagger$,
it contributes only to odd-point amplitudes.

\medskip
\mysec{Positivity from Cauchy-Schwarz}The logic of our multipositivity constraints is simple.   To begin, consider the residue of an $(n+m)$-point amplitude $A^{(n+m)}$ on a 
factorization channel.  Mechanically, $A^{(n+m)}$ exhibits a simple pole at $s=M^2_k$, where $s$ is the kinematic invariant corresponding to the channel and $M_k$ is the mass of the exchanged state labeled by $k$.   The residue on this pole factorizes into a product of lower-point amplitudes,
\be
\begin{aligned}
&  R^{(n+m)}_{k}(p_1,\!... ,p_n|q_{1},\!...,q_m) \\
& \ \ \ \ \ \ \ = A^{(n+1)}_{k}  (p_1,\!..., p_n) A^{(m+1)}_{k}(q_1,\!..., q_m),
\label{eq:fact_fg}
\end{aligned}
\ee
where $s \equiv -(p_1+ \cdotn + p_n)^2=-(q_1+ \cdotn + q_m)^2$, with all momenta incoming. Since all these amplitudes are planar, their momenta  $p_i$ and $q_i$ are cyclically ordered. 
  If there is a degeneracy of states at mass $M_k$, then the right-hand side is augmented to a sum that will not affect our conclusions. 
Next, consider the residues of a $2n$-point amplitude $A^{(2n)}$ and a $2m$-point amplitude $A^{(2m)}$,
\be \hspace{-2mm}
\begin{aligned}
R^{(2n)}_{k}(p_1,\!... ,p_n| {-}\bar p_n,\!...,{-}\bar p_1) &=|A^{(n+1)}_{k}(p_1,\!...,p_n)|^2 \\
R^{(2m)}_{k}({-}\bar q_m,\!...,{-}\bar q_1 | q_1,\!...,q_m) &= |{A}^{(m+1)}_{k}(q_1,\!...,q_m)|^2 ,
\end{aligned}\hspace{-1mm}\label{eq:Rpos}
\ee
where bar denotes complex conjugation. We  have chosen complex forward kinematics~\cite{Cheung:2025nhw}, making the residues modulus squares since $A^{(n+1)}_k(-\bar p_1,\!...,-\bar p_n)=\bar A^{(n+1)}_k(p_1,\!...,p_n) $. 

Recasting the lower-point amplitudes as the components of a vector, $\bar A^{(n+1)}_{k} \equiv\langle k |f \rangle  M_k^{i+1}  $ and $A^{(m+1)}_{k} \equiv   \langle k |g \rangle M_k^{j+1} $, 
where for future convenience we normalize with different powers of the masses,  we find that
\eq{
R^{(2n)}_{k}& =  \langle f| k\rangle \langle k |f\rangle M_k^{2i+2}  \\
R^{(2m)}_{k} &=   \langle g| k\rangle \langle k |g\rangle M_k^{2j+2} \\
R^{(n+m)}_{k} &= \langle f| k\rangle \langle k |g\rangle M_k^{(i+j)+2} .
}{eq:R_list}
Here $|f\rangle$ and  $|g\rangle$  are $n$- and $m$-particle states with momenta $p_{1},\cdotn, p_n$ and $q_{1}, \cdotn , q_m$, respectively, while $|k\rangle$ is the one-particle exchanged state of mass $M_k$, weighted by any degeneracies.

At low energies, we expand the amplitudes $A^{(r)}$ in terms of Wilson coefficients associated with powers of $s$:
\eq{
A^{(r)}\to \sum_\ell \lambda^{(r)}_{\ell}s^\ell, \quad \textrm{where} \quad \lambda^{(r)}_{\ell} =\underset{s=0}{\text{Res}}\left(\frac{A^{(r)}}{s^{\ell+1}}\right).
}{}
 Note that the Wilson coefficients $ \lambda^{(r)}_{\ell}$ depend implicitly on all kinematic invariants that are independent of $s$.
 Applying  Cauchy's theorem  to the integral $\oint ds\,A^{(r)}/s^{\ell+1}$ taken over
a   contour  at infinity in the complex $s$-plane, with  all other kinematic invariants  fixed, for $(r,\ell)=(2n,i),(2m,j),(n+m,(i+j)/2)$,
we obtain
\be \hspace{-2mm}
\begin{aligned}
\lambda^{(2n)}_{i} &= \underset{s=0}{\text{Res}}\left(\!\frac{A^{(2n)}}{s^{i+1}}\!\right) =\sum_k \frac{R^{(2n)}_{k}}{M_k^{2i+2}}  = \langle f|f\rangle ,\\
\lambda^{(2m)}_{j} &= \underset{s=0}{\text{Res}}\left(\!\frac{A^{(2m)}}{s^{j+1}}\!\right)  =\sum_k \frac{R^{(2m)}_{k}}{M_k^{2j+2}}  = \langle g|g\rangle ,\\
\lambda^{(n+m)}_{(i+j)/2} &=\underset{s=0}{\text{Res}}\left(\!\frac{A^{(n+m)}}{s^{(i+j)/2+1}}\!\right) =\sum_k \frac{R^{(n+m)}_{k}}{M_k^{(i+j)+2}}  =  \langle f|g\rangle .\label{eq:rewrite_as_vectors}
\end{aligned}\hspace{-1mm}
\ee
Here, the planarity of the amplitudes  was crucial  to relate the residue in $s=0$ to the sum of residues across a {\it single} factorization channel. 
We have  assumed that every subtracted amplitude has sufficiently convergent high-energy behavior such that we can drop boundary terms. We will comment later on particular choices of $i,j$ such that this assumption is well motivated.

Applying the Cauchy-Schwarz inequality $|\langle f|g\rangle |^2 \leq \langle f|f\rangle \langle g|g\rangle$ to \Eq{eq:rewrite_as_vectors}, we find that
\begin{equation}
|\lambda^{(n+m)}_{(i+j)/2}|^2 \leq \lambda^{(2n)}_{i} \lambda^{(2m)}_{j}.
\label{eq:CS-res}
\end{equation} 
The constraint above implicitly depends on a set of $s$-independent kinematic invariants.

To project \Eq{eq:CS-res} into a particular  $s$-independent kinematic invariant, 
it is convenient to introduce the following ansatz for the parameterization of momenta
\be
p_i \equiv p_i(\{a\},s) \text{ and } q_i \equiv q_i(\{b\},s), 
\label{para}
\ee
where $\{a\}$ and $\{b\}$ are sets of parameters that we will define explicitly in the next section.
Rerunning  the above logic, identifying $\partial_{a}  A^{(n+1)}_{k} $ and $\partial_{b}  A^{(m+1)}_{k}$ as the vector components in the Cauchy-Schwarz inequality,  we find
\begin{equation}
|\partial_{a}\partial_{b} \lambda^{(n+m)}_{(i+j)/2}|^2 \leq \partial_{a}\partial_{\bar a}\lambda^{(2n)}_{i} \partial_{b}\partial_{\bar b}\lambda^{(2m)}_{j}.
\label{eq:CS-res-deriv}
\end{equation} 
Factorization of the derivatives requires that the left cluster only depend on the momenta $p_i$ and therefore on the parameters $\{a\}$, while the right cluster only on the momenta $q_i$ and therefore on the parameters $\{b\}$.
This holds to all derivative orders. 
The simplest version of  \Eq{eq:CS-res}  has $n=2$ and $m=3$, 
as shown in Fig.~\ref{fig:456bound}, giving
\eq{
| \lambda^{(5)}_{(i+j)/2}|^2 &\leq \lambda^{(4)}_{i} \lambda^{(6)}_{j} ,\\
|\partial_a \partial_b \lambda^{(5)}_{(i+j)/2}|^2 &\leq  \partial_a\partial_{\bar a} \lambda^{(4)}_{i} \partial_b \partial_{\bar b} \lambda^{(6)}_{j},
}{eq:456-bound} 
which constrains the Wilson coefficients of the four-, five-, and six-point amplitudes.
We will focus on this system in the rest of the paper and study the implications of \Eq{eq:456-bound} for the chiral Lagrangian.  

\medskip

\mysec{Parameterization of Momenta}The bounds we have derived are true only provided that we can construct the states promised in Eq.~\eqref{eq:R_list}, which allow us to write the dispersion relations in terms of inner products in Eq.~\eqref{eq:rewrite_as_vectors}. 
We now fulfill this promise.
For these vectors to be well defined, we must specify on-shell external momenta $p_i$ and $q_i$ in terms of $s$ and a handful of other parameters as in Eq.~\eqref{para} such that the four-, five-, and six-point amplitudes are analytic in $s$ and obey momentum conservation.
For  \Eq{eq:CS-res}  to hold, we want $s= -(p_1+\cdotn + p_n)^2$ to activate singularities in a single factorization channel.   

The three amplitudes relevant for our multipositivity bounds are $A^{(2n)}(p_1,\!...,p_{2n})$, $A^{(2m)}(q_1,\!...,q_{2m})$ and $A^{(n+m)}(p_1,\!...,p_n,q_1,\!...,q_m)$. To obtain \Eq{eq:rewrite_as_vectors}, we need the residues of $A^{(2n)}$ and $A^{(2m)}$ on the factorization channel to be positive. This requires the complex forward limit of Eq.~\eqref{eq:Rpos}, $(p_{n+1},\!..., p_{2n})\to (-\bar  p_n, \!..., -\bar  p_1)$ and $(q_{m+1}, \!..., q_{2m}) \to(-\bar q_{m}, \!..., -\bar q_1)$, for which momentum conservation imposes a constraint,
 \be 
\begin{aligned}
(\bar  p_1 + \cdotn + \bar  p_n)&=(p_1+ \cdotn +p_n),\\
(\bar q_{1}+ \cdotn + \bar q_m)&=(q_1+ \cdotn +q_m).
\label{eq:5pt-cluster-reality}
\end{aligned}
\ee

We are now ready to construct explicit momenta for the system comprising four-, five-, and six-point scattering.
It will be convenient to introduce a tetrad basis,
\be
\begin{aligned}
\ell_\pm^\mu&=\frac{1}{2}(1,0,0,\pm 1),\;\;\; &
n_\pm^\mu&=\frac{1}{2}(0,1,\pm i,0),
\end{aligned}
\label{eq:tetrad}
\ee
where the only nonzero contractions are
$\ell_+\cdot\ell_-=-1/2$ and $
n_+\cdot n_-=1/2$. Following \Eq{para} we define
\be
\begin{aligned}
p_1&=\sqrt{s}\, \ell_+ + a n_+  \\
p_2&=\sqrt{s}\, \ell_- - a n_+   \\
q_1&=-\sqrt{s}\, \ell_- + b_1 n_-  \\
q_2&=-b_1 b_3/\sqrt{ s}\,  \ell_+ -(b_1-b_2)n_-   \\
q_3&=(b_1 b_3/\sqrt{ s}-\sqrt{s})\, \ell_+ - b_2 n_- ,   \\
\end{aligned}
\label{eq:5pt-generic-ansatz}
\ee
where $a$ and $b_{1,2,3}$ are complex numbers. Notice that by construction $(p_1+p_2)\propto(q_1+q_2+q_3)\propto \sqrt{s}$. Momentum conservation in \Eq{eq:5pt-cluster-reality} requires real $s$ on the factorization channel in Eq.~\eqref{eq:rewrite_as_vectors}, where $s=M_k^2$~\footnote{Note that we only need to invoke an explicit parameterization of the momenta {\it after} localizing the contour integral to a sum over physical resonances.  In particular, the parameterization applies {\it on the locus of any factorization channel}, where $s$ is real and set to $M_k^2$ on each branch of $|k\rangle\langle k|$, where in Eq.~\eqref{eq:rewrite_as_vectors} we have used $\protect\sum_k |k\rangle\langle k|$ to define the inner product.}. 

Plugging these momenta into the kinematic invariants of each amplitude, we find that 
for $A^{(4)}(p_1,p_2,-\bar p_2,-\bar p_1)$,
\be 
-(p_1+p_2)^2= s,\qquad -(p_2 - \bar p_2)^2=a \bar{a}, 
\label{eq:4ptmands}
\ee
while for $A^{(5)}(p_1,p_2,q_1,q_2,q_3 )$,
\be 
\begin{aligned}
-(p_1+p_2)^2&= s,& -(p_2+q_1)^2&= a b_1 ,  \\
-(q_1+q_2)^2&= b_1 b_3, &-(q_2+q_3)^2&= 0, \\
-(q_3+p_1)^2&= a b_2,
\label{eq:5ptmands}
\end{aligned}
\ee
and for $A^{(6)}(-\bar q_3,-\bar q_2,-\bar q_1,q_1,q_2,q_3)$,
\be  \hspace{-3mm}
\begin{aligned}
-(\bar{q}_3+\bar{q}_2)^2&=0, &-(\bar{q}_2+\bar{q}_1)^2&=\bar{b}_1\bar{b}_3 , \\ -(\bar{q}_1-q_1)^2&=b_1\bar{b}_1,  
& -(q_1+q_2)^2&=b_1 b_3, \\
-(q_2+q_3)^2&= 0 , &-(q_3-\bar{q}_3)^2&= b_2 \bar{b}_2, \\
-(\bar{q}_2+\bar{q}_1-q_1)^2&= b_1 \bar{b}_2, &-(\bar{q}_1-q_1-q_2)^2&=b_2\bar{b}_1, \\
-(q_1+q_2+q_3)^2&= s.
\label{eq:6ptmands}
\end{aligned}
\ee

One might have expected the five-point amplitude to depend on five independent Mandelstam invariants. However, as visible in \Eq{eq:5ptmands}, the counting is reduced by one. This happens because the low-energy five-point amplitude coming from the WZW term 
$\propto\epsilon_{\mu\nu\rho\sigma}\,p_1^\mu p_2^\nu q_1^\rho q_2^\sigma$,
when expressed in terms of Mandelstam invariants, is proportional to the square root of $A_5 s^2+B_5 s+C_5$, where $A_5$, $B_5$, and $C_5$ are functions of the other kinematic invariants.  
Requiring the amplitude to be analytic in $s$ imposes an additional tuning among the other Mandelstams, $B_5=\pm 2\sqrt{A_5 C_5}$ , which brings the counting of independent Mandelstams down to four, as given in \Eq{eq:5ptmands}. A similar situation arises at six points, where we would expect nine independent Mandelstams. In four dimensions, however, we have a Gram constraint, which in principle reduces the counting by one. The Gram constraint can be written as $A_6 s^2+B_6 s+C_6=0$, where $A_6$, $B_6$, and $C_6$ are functions of the other kinematic invariants. 
Satisfying the Gram constraint while  keeping  the other invariants $s$-independent---so that a single channel is activated---requires  $A_6=B_6=C_6=0$, reducing the counting of free parameters to six, as visible in \Eq{eq:6ptmands}.

\medskip

\mysec{A Higher-Point Bound in Large-$N_c$ QCD}We are equipped with a kinematic chart that describes the amplitudes $A^{(4)}$, $A^{(5)}$, $A^{(6)}$ in terms of the same set of variables, in particular allowing us to write their residue sums as inner products as in Eq.~\eqref{eq:rewrite_as_vectors} and thereby giving us the multipositivity bound in Eq.~\eqref{eq:456-bound} . 
By explicit calculation, the planar partial  amplitudes---as prescribed in \Eq{nptplanar}---associated with the EFT Lagrangian for the NLSM in Eqs.~\eqref{chirallagrangian} and \eqref{WZWpion} are
\begin{widetext}
\eq{
A^{(4)}&= \frac{s+|a|^2}{2F_\pi^2} 
+\frac{2(L_3+2L_4)(s^2+|a|^4)+8L_4s |a|^2}{F_\pi^4}+ \cdotn , \hspace{1cm} A^{(5)} = \frac{i \kappa}{24\sqrt{2} F_\pi^5 \pi^2} a \left[s(b_1-b_2)-b_1^2b_3\right]+ \cdotn , \\
A^{(6)} &= \frac{|b_1-b_2|^2+ (b_1 b_3 - b_1 \bar b_2 b_3/\bar b_1 +\mathrm{c.c.})}{4F_\pi^4}
- \frac{|b_1|^2 |b_3|^2}{4F_\pi^4 s}+\frac{(L_3+2L_4)}{F_\pi^6 |b_1|^2 }\left[|b_1|^2\left(|b_1|^2+|b_2|^2\right)^2
+ |b_1|^4 |b_3|^2\right]\\& - \frac{(L_3+2L_4)}{F_\pi^6 |b_1|^2 }
\left[
|b_1|^2(|b_1|^2{+}|b_2|^2)\bar{b}_1 b_2
- b_1^2 b_3(\bar{b}_1{-}\bar{b}_2)
  (|b_2|^2{+}b_1 b_3){+} \frac{|b_1|^4 |b_3|^2 b_1 b_3}{s}
{+} \text{c.c.}
\right]+\frac{4|b_1{-}b_2|^2 L_4 s}{F_\pi^6}+\cdotn, }{}
\end{widetext}
where ellipses indicate terms of higher order in the derivative expansion.
We now want to determine the number of subtractions we should take, i.e., the choices of  $i$ and $j$ that make \Eq{eq:CS-res} and later inequalities true. To do so, we must consider the high-energy limit of the four-, five-, and six-point amplitudes with the other kinematic parameters fixed.
The Regge behavior of higher-point amplitudes is highly nontrivial and has not been studied with the same rigor as for $2 \to 2$ processes~\footnote{Even in the case of open-string amplitudes at tree level, the detailed Regge structure at higher multiplicity was only recently investigated in Ref.~\cite{Arkani-Hamed:2024nzc}.}. 
The single contour integration in $s$ means that to drop the boundary term we have to study the behavior of $A^{(n)}$ as $s\to\infty$ with all other kinematic invariants kept fixed. In this limit, planar Reggeized amplitudes at tree level obey the structure $A^{(n)} \rightarrow s^{{\rm max}\{\alpha_1,\alpha_2,\ldots,\alpha_{n-3}\}}$,
where $\alpha_1,\!...,\alpha_{n-3}$ correspond to the intercepts of the Regge trajectories along cuts of a half-ladder diagram~\cite{Brower:1974yv,Collins:1977jy}.
Large-$N_c$ QCD does not have a pomeron in the leading trajectories, so all $\alpha_i<1$ and $A^{(4)}$, $A^{(5)}$, and $A^{(6)}$ all grow more slowly than $s$. A once-subtracted dispersion relation is therefore sufficient to drop the boundary term, i.e., we can take $i,j\geq 1$ in  Eqs.~\eqref{eq:CS-res} and \eqref{eq:CS-res-deriv}.
 
We are interested in the  Wilson coefficients linear in $s$:
\begin{equation}
\begin{aligned}
\lambda^{(4)}_1 &= \frac{1}{2\Fpi^2}+\frac{8L_4}{\Fpi^4} |a|^2+\cdots \\
\lambda^{(5)}_1 &= \frac{i \kappa a(b_1-b_2)}{24\sqrt{2} \Fpi^5 \pi^2}+\cdots\\ 
\lambda^{(6)}_1 &= \frac{4L_4|b_1-b_2|^2}{\Fpi^6}+\cdots .
\end{aligned}
\end{equation}
To eliminate contributions from yet higher-derivative terms and ensure that only terms quartic in momenta  appear in our bounds, we differentiate with respect to $s$-independent parameters in Eq.~\eqref{eq:5pt-generic-ansatz} and then pin the parameters to zero,
\be
|\partial_a \partial_{b_1} \lambda^{(5)}_{1}|_{a,b_i=0}^2 \leq  \partial_a\partial_{\bar a} \lambda^{(4)}_{1} \partial_{b_1} \partial_{\bar b_1} \lambda^{(6)}_{1}|_{a,b_i=0},
\ee
which leads to
\begin{equation}
\left| \frac{i \kappa}{24\sqrt{2} \Fpi^5 \pi^2}\right|^2 \leq\left(\frac{8L_4}{\Fpi^4}\right)\left(\frac{4L_4}{\Fpi^6}\right).
\label{eq:boundLs}
\end{equation}
Recall that in large-$N_c$ QCD, the parameters scale as $F_\pi^2 \propto L_4 \propto N_c$ and $\kappa = N_c$ , so $N_c$ cancels on both sides.

\begin{figure}[t]
\includegraphics[width=\columnwidth]{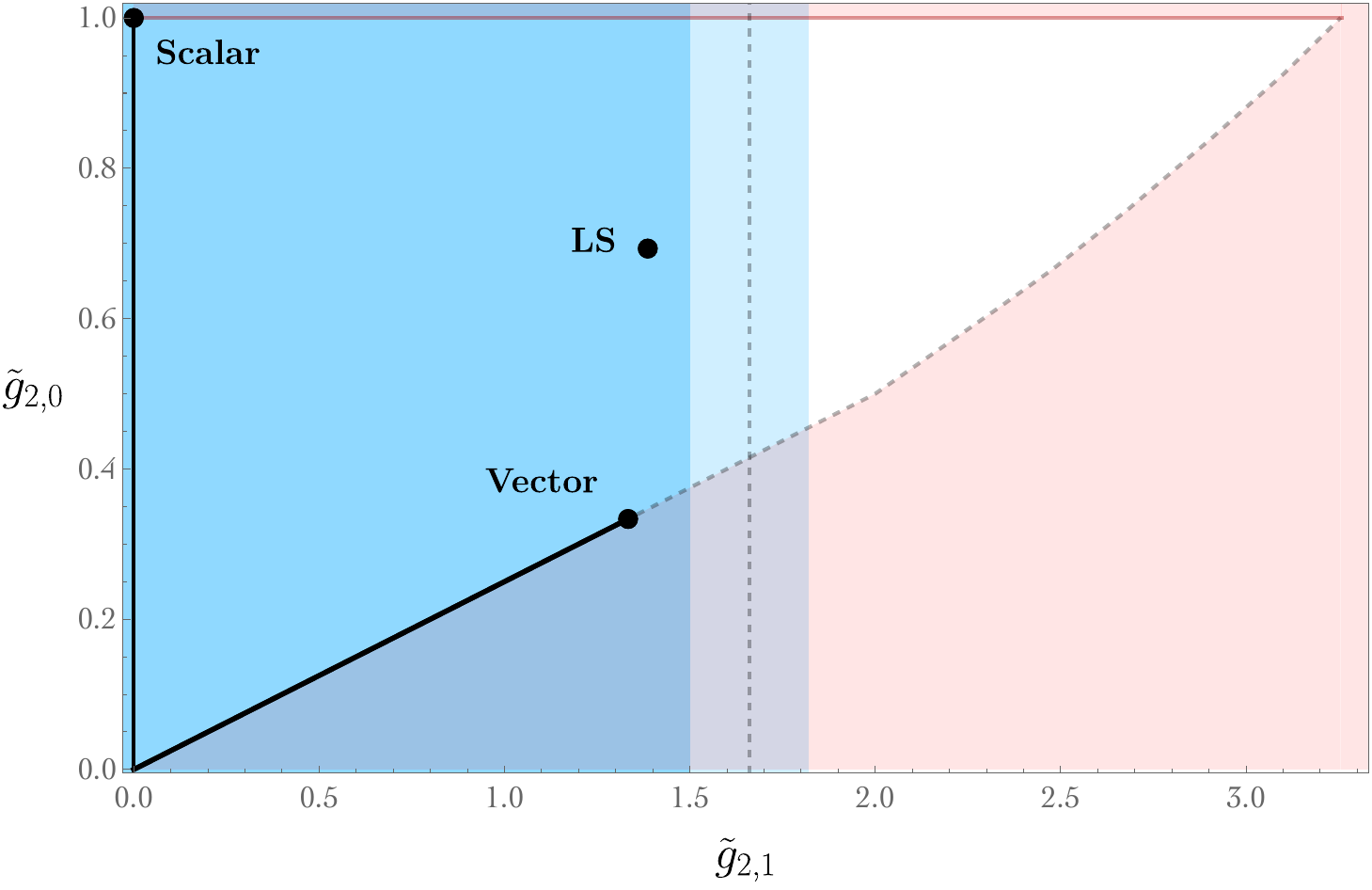}
\caption{Multipositivity bound in \Eq{eq:g21tildebound}
for large-$N_c$ QCD (in blue) overlaid  on the exclusion plot derived from four-point positivity~\cite{Albert:2022oes, Fernandez:2022kzi} (in red). 
The band delimiting the exclusion area represents the numerical uncertainty in 
 the lattice computations~\cite{Bali:2013kia}. }
\label{fig:largeNbound}
\end{figure}

To confront this new constraint with bounds previously obtained from four-point positivity, let us define the dimensionless and $N_c$-independent quantities $\tilde{g}_{2,1}=16L_4M^2/F_\pi^2$ and $\tilde{g}_{2,0}=4(L_3+2L_4)M^2/F_\pi^2$,
where $M$ is the mass of the lightest meson exchanged in $\pi\pi$ scattering~\cite{Albert:2022oes, Fernandez:2022kzi} and $L_4 \geq 0$.
The bound in \Eq{eq:boundLs}  becomes
\begin{equation}
\tilde g_{2,1}\geq \frac{N_c M^2}{12\pi^2  \Fpi^2} \simeq 1.66 ,
\label{eq:g21tildebound}
\end{equation}
where in the last relation we have used large-$N_c$  lattice computations~\cite{Bonanno:2025hzr} to fix the $\rho$ meson mass $M \simeq 8.1 \sqrt{3/N_c} \Fpi$~\footnote{Here we used the ratio $M_\rho \sqrt{N_c}/F_\pi$ extracted from Ref.~\cite{Bonanno:2025hzr}.  The ratio was also computed in an older lattice analysis~\cite{Bali:2013kia}, which leads to a $\sim 25\%$ weaker bound. This earlier computation, however, was performed at fixed lattice spacing, while Ref.~\cite{Bonanno:2025hzr} extrapolates to the continuum. For other large-$N_c$ QCD studies on the lattice, see also Ref.~\cite{Hernandez:2020tbc}}.
This leads to a new exclusion region highlighted in blue in Fig.~\ref{fig:largeNbound}.  
One sees that the  bound severely constrains the  scalar contribution to $\tilde{g}_{2,1}$, and is marginally inconsistent with a model where a single vector is exchanged, as well as the stringy Lovelace-Shapiro amplitude (LS), subject to lattice uncertainties.

\medskip

\mysec{Discussion}In this paper, we have derived new constraints on the chiral Lagrangian. At the heart of our analysis is the insight that certain multiparticle scattering processes are mathematically inconsistent.  Our results are the first phenomenologically relevant application of this fact. Using the four-, five-, and six-point pion amplitudes of the chiral Lagrangian, we deduced a novel bound on the size of the leading higher-derivative interactions with respect to the WZW term, yielding a remarkable mutual constraint on anomalous and nonanomalous physics. Previous attempts~\cite{Albert:2023jtd, Ma:2023vgc} to bound the chiral anomaly relied on, and were therefore limited by, four-point amplitudes, requiring photons as external particles and thus introducing an additional sector. Our methods apply to tree-level planar amplitudes, which are the natural setting for large-$N_c$ QCD.

The present analysis leaves numerous avenues for future exploration. Even in the case of planar tree-level amplitudes, we have only scratched the surface by constraining operators that are leading order in the momentum expansion.  Our methods should generalize mechanically to the full  tower of higher-derivative interactions in the chiral Lagrangian.   While ever more operators will of course appear, with corresponding unknown Wilson coefficients, it may be possible to choose kinematic limits to tame this growth of free parameters. This work opens the door for a systematic numerical study~\cite{Simmons-Duffin:2015qma} of higher-point constraints.

Since we are analytically continuing in a single Mandelstam variable, it is tempting to generalize beyond tree level. Indeed, if a Lehmann-K\"all\'en representation holds on a single cut, then our results immediately hold nonperturbatively, provided that the higher-point amplitude obeys analyticity in $s$ and our Regge boundedness conditions. However, the analytic structure of higher-point amplitudes has not been as axiomatically investigated as the four-point case and is famously subtle, so caution is warranted in extending our results nonperturbatively. 

A more technically fraught direction is nonplanar scattering, relevant to finite-$N_c$ QCD. Our construction hinges on detailed control over which factorization channels are activated as the center-of-mass energy is scanned. In nonplanar amplitudes, additional channels and multi-trace structures can appear, so it remains to be seen whether these methods extend in a useful form.
We leave this question to future investigation.

\medskip

\noindent {\it Acknowledgments:} 
We thank Fernando Romero-L\'{o}pez for discussions.
C.C. is supported by the Department of Energy (Grant No.~DE-SC0011632), the Walter Burke Institute for Theoretical Physics, and the Leinweber Forum for Theoretical Physics.  J.J.~and P.K.~have been supported by KIAS Individual Grants (QP090001) [J.J.] and (QP021402) [P.K.] through the Quantum Universe Center at the Korea Institute for Advanced Study, and P.K.~has also been supported by the National Research Foundation of Korea (NRF) funded by the Ministry of Education, Science and Technology (RS-2025-24803289). A.P. is supported by the ICREA Acad\`emia program. A.P. and F.S. are
supported by the research grants 2021-SGR-00649, PID2023-146686NB-C31, and CEX2024-001441-S. F.S is also supported by the grant PRTR-C17.I1.
G.N.R. is supported by the James Arthur Postdoctoral Fellowship at New York University.

\bibliographystyle{utphys-modified}
\bibliography{multipositivity}

\bigskip

\appendix

\newcommand{\J}{\mathcal{J}}

\mysec{Appendix}For the sake of completeness, we will briefly summarize our calculation of amplitudes in chiral perturbation theory.  
We employ the alternative formulation derived in Ref.~\cite{Slavnov:1971mz}, which is highly efficient for our purposes. This formulation was later shown to encode color-kinematics duality directly in terms of fields and currents~\cite{Cheung:2021zvb}. In this approach, all dynamics are encoded in the chiral current,
\begin{equation}
j_\mu = i\,U^\dagger \partial_\mu U,
\end{equation}
which is independent of the choice of embedding of the pion field within $U$.

The action governing chiral perturbation theory is
\begin{equation}
S=S_{\rm NLSM}+S_{\rm WZW},
\end{equation}
where the first and second terms encode the parity-even nonanomalous and parity-odd anomalous dynamics, respectively.
In terms of the chiral current, Eq.~\eqref{chirallagrangian} becomes
\begin{equation}\hspace{-2mm}
    \mathcal{L}_{\rm NLSM}
   \,{=}\, {-}\frac{\Fpi^2}{4}\,{\rm Tr}(j^2)
    {+} L_3{\rm Tr}(j^2j^2)
    {+} L_4{\rm Tr}(j_\mu j_\nu j^\mu j^\nu) .
    \hspace{-1mm}
\end{equation}
Since $j_\mu$ takes the form of a pure gauge configuration, its field strength vanishes identically,
\begin{equation}
    F_{\mu\nu}
    = \partial_\mu j_\nu - \partial_\nu j_\mu - i[j_\mu,j_\nu]
    = 0 .
    \label{eq:flatness}
\end{equation}
At linearized order, the chiral current is
\begin{equation}
j_\mu
=
-\frac{\sqrt{2}}{\Fpi}\partial_\mu\pi
+O(\pi^2),
\label{eq:j-linearized}
\end{equation}
which will be important later. Note also that
\begin{equation}
-\frac{\Fpi^2}{4}
{\rm Tr}(j_\mu j^\mu)
=
-\frac12
\partial_\mu\pi^a
\partial^\mu\pi^a
+O(\pi^3),
\end{equation}
so the pion kinetic term is canonically normalized.

To incorporate the WZW action, we specify the five-dimensional manifold ${\cal M}_5={\cal M}_4\times [0,1]$ and take 
$U(x;y)$ to depend on an auxiliary coordinate $y$ such that the ordinary chiral field corresponds to $y=1$.

To derive the equations of motion, we apply an infinitesimal variation $U+\delta U
=
Ue^{-i\,\delta\phi}$, 
so that $\delta U=-i\,U\delta\phi$ and $\delta U^{-1}=i\,\delta\phi\,U^{-1}$.
The chiral current varies by
\begin{align}
\delta j_\mu
&=
i(\delta U^{-1})\partial_\mu U
+iU^{-1}\partial_\mu(\delta U)
\nonumber\\
&=
\partial_\mu\delta\phi
-i[j_\mu,\delta\phi]
=
D_\mu\delta\phi ,
\label{eq:delta-j-deltaphi}
\end{align}
which defines the covariant derivative.  Integrating by parts, we obtain the variation of the NLSM action,
\begin{equation}
\delta S_{\rm NLSM}
=
\int d^4x\,
\frac{\Fpi^2}{2}\,
{\rm Tr}\!\left(
\delta\phi\,
D_\mu \J^\mu
\right),
\label{eq:deltaL}
\end{equation}
where we have defined
\eq{
\J^\mu
&=
j^\mu
-\frac{4L_3}{\Fpi^2}
\left(
j^\mu j^2+j^2j^\mu
\right)
-\frac{8L_4}{\Fpi^2}
j^\nu j^\mu j_\nu.
}{eq:J-def}
The identity $F_{\mu\nu}=0$ implies
\begin{equation}
D_\mu\J^\mu
=
\partial_\mu\J^\mu,
\end{equation}
which reduces to the conservation law $\partial_\mu\J^\mu=0$ in the absence of the WZW action.

Meanwhile, the variation of the WZW action is 
\be \hspace{-2mm}
\delta S_{\rm WZW}{=}{-}\frac{\kappa}{48\pi^2}\!\!
\int \!\!d^4x\!\!\int_0^1\!\!\!dy\,
\epsilon_{\tau\mu\nu\rho\sigma}
{\rm Tr}\!\left(
(D^\tau\! \delta\phi)
j^\mu \! j^\nu \! j^\rho \! j^\sigma\!
\right)\! .\hspace{-1mm}
\ee 
To strip off the covariant derivative acting on $\delta\phi$, we use
\begin{equation}
\epsilon_{\tau\mu\nu\rho\sigma}
\,{\rm Tr}\!\left[
\delta\phi\,
D^\tau
(j^\mu j^\nu j^\rho j^\sigma)
\right]
=
0,
\end{equation}
which follows from the antisymmetry of the Levi-Civita tensor together with $F_{\mu\nu}=0$. Then we have
\be  \hspace{-1.5mm}
\delta S_{\rm WZW}
=-
\frac{\kappa}{48\pi^2}
\int d^4x\,
\epsilon_{\alpha\beta\rho\sigma}
{\rm Tr}\!\left(
\delta\phi\,
j^\alpha j^\beta j^\rho j^\sigma
\right)\!\Big|_{y=1}, \hspace{-1mm}
\ee
which reduces the WZW variation to a four-dimensional chiral current interaction.

Combining both contributions from the NLSM and WZW actions at $y=1$ yields
\begin{align}
\frac{\Fpi^2}{2}\partial_\mu\J^\mu
-
\frac{\kappa}{48\pi^2}
\epsilon_{\alpha\beta\rho\sigma}
j^\alpha j^\beta j^\rho j^\sigma
=
0,
\end{align}
which we rewrite as
\be 
\begin{aligned}
\partial_\mu j^\mu
&=
\frac{1}{\Fpi^2}
\partial^\mu\!\Bigl[
4L_3
\bigl(
j_\mu j^2+j^2j_\mu
\bigr)
+
8L_4
j_\nu j_\mu j^\nu
\Bigr]
\\
&\quad
+\frac{\kappa}{24\pi^2\Fpi^2}
\epsilon_{\alpha\beta\rho\sigma}
j^\alpha j^\beta j^\rho j^\sigma ,
\label{eq:divergence-eom-rearranged}
\end{aligned}
\ee
which is a modified conservation law for the chiral current.

We now reinterpret this equation as a Berends-Giele recursion relation for amplitudes in chiral perturbation theory. To do so, we introduce a source $J$ for the pion,
\eq{
\Box\pi
&=
J+O(\pi^2).
}{}
In the presence of this source, Eq.~\eqref{eq:j-linearized} gives
\begin{equation}
\partial_\mu j^\mu
=
-\frac{\sqrt{2}}{\Fpi}J
+O(\pi^2).
\end{equation}
Following Ref.~\cite{Cheung:2021zvb}, we compute the divergence of the flat field strength $\partial^\mu F_{\mu\nu}=0$ to obtain
\begin{equation}
\Box j_\nu
-\partial_\nu(\partial_\mu j^\mu)
-i\partial^\mu[j_\mu,j_\nu]
=
0 .
\label{eq:current-second-order-pre}
\end{equation}
Then, substituting Eq.~\eqref{eq:divergence-eom-rearranged} yields
\begin{equation}\hspace{-2mm}
\begin{aligned}
-\frac{\sqrt{2}}{\Fpi}\partial_\nu J&=
\Box j_\nu
-i\partial^\mu[j_\mu,j_\nu]
\\
&
-\frac{\partial_\nu\partial^\mu}{\Fpi^2}
\left[
4L_3
\left(
j_\mu j^2+j^2j_\mu
\right)
+
8L_4
j_\rho j_\mu j^\rho
\right]
\\
&
-\frac{\kappa}{24\pi^2\Fpi^2}
\partial_\nu\!
\left(
\epsilon_{\alpha\beta\rho\sigma}
j^\alpha j^\beta j^\rho j^\sigma
\right).
\end{aligned}
\label{eq:current-second-order}
\end{equation}
Since $\partial^\mu F_{\mu\nu}=0$ is purely geometric and independent of the action, all higher-order interactions enter only through the deformation of the current divergence.  
Note that the pion source acts as an external polarization vector for the chiral current, $\epsilon_\mu=-\sqrt{2}ip_\mu/\Fpi$.  

By iteratively solving this equation, we derive the one-point function of the chiral current at arbitrary perturbative order. The resulting recursive structure is considerably simpler than the standard pion Lagrangian expansion.  To compute the $n$-point scattering amplitude, we compute the one-point function of the chiral current with $n-1$ pion sources, each entering with its own polarization $\epsilon_\mu$.

The above construction computes the one-point function of the chiral current rather than the pion field itself. However, the two are related at linearized order by
\eq{
\pi
&=
-\frac{\Fpi}{\sqrt{2}}
(q^\mu\partial_\mu)^{-1}
q_\mu j^\mu
+O(\pi^2),
}{}
where $q$ is an arbitrary reference vector and $(q^\mu\partial_\mu)^{-1}$ is defined mechanically in momentum space. To extract the pion one-point function, we take the momentum on shell and choose the polarization vector, $\tilde\epsilon_\mu=i\Fpi\,q_\mu/\sqrt{2}\,qp $.  By dotting the one-point function of the chiral current with $n-1$ pion sources into the final polarization $\tilde \epsilon_\mu$, we obtain the $n$-point pion amplitude.

The explicit amplitudes from our calculation can be found in the supplemental files.

\end{document}